\documentclass[12pt,twoside]{article}
\usepackage{fleqn,espcrc1}
\usepackage{graphicx,epsfig}

\newcommand{\AmS}{{\protect\the\textfont2
  A\kern-.1667em\lower.5ex\hbox{M}\kern-.125emS}}

\title{The photoproduction of vector mesons }

\author{J.-M. Laget \address{CEA-Saclay, DAPNIA-SPhN, 
        F91191 Gif-sur-Yvette, France}
     \thanks{Work supported by French Atomic Energy Commission and partly
        by European Commission under Contract HPRN-CT-2000-00130.}
}
       
\begin{document}

\maketitle

\begin{abstract}
At high energy, the photoproduction or electroproduction of Vector Mesons allow to prepare a beam of quark-antiquark pairs of a given flavor. At high momentum transfer, the study of the scattering  of these pairs on a nucleon opens up an original window on the quark-gluon structure of hadronic matter, which may eventually shed light on  its gluonic content, on correlations between quarks and on the van der Walls part of the interaction between hadrons.  
\end{abstract}

\section{INTRODUCTION}

Since a photon has the same quantum numbers as vector mesons, a beam of high energy photons is a beam of quark-antiquark pairs. Their flavor is selected by the kind of vector meson which is detected: light quarks for $\rho$ or $\omega$, strange quarks for $\phi$ and charmed quarks for $J/\psi$ mesons. The life time of the fluctuation $2\nu /(Q^2+M_V^2)$ is
given by the uncertainty principle and increases with the beam energy $\nu$: for instance, a 5 GeV beam of real photons fluctuates into a beam of $\rho$ mesons over 4 fm. Therefore, the interaction with matter of a beam of energetic photons is similar to the interaction of hadrons.

\begin{figure}[h]
\centerline{\epsfig{file=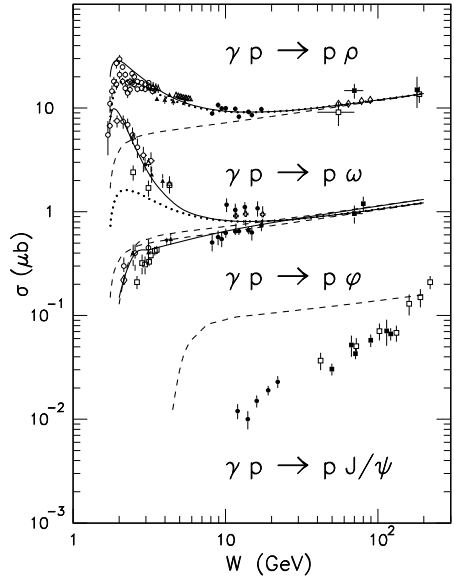,height=3.0in}}
\vspace{-1cm}
\caption{The total cross-sections of the various vector meson photoproduction channels. Dashed lines: Pomeron exchange. Dotted lines include also $f_2$ meson exchange.  Full lines include in addition $\pi$ and $\sigma$ exchanges.}
\label{vect_tot}
\end{figure}

This is illustrated in Fig.~\ref{vect_tot}, which shows the variation, against the available c.m. energy $W=\sqrt{s}$, of the cross section of  the photoproduction of vector mesons~\cite{Dur00}, from threshold up to the HERA energy range. The rise with energy is due to the Pomeron exchange ($\propto s^{0.16}$), while the exchange of the $f_2$ Regge trajectory ($s^{-1}$) is necessary to reproduce the trend of the cross section in the intermediate energy range of SLAC and FNAL. At low energy, the exchange of the Regge trajectories of the $\sigma$ and $\pi$ mesons account  for the $s^{-2}$ fall off of the cross section in the $\rho$ and  $\omega $ channels respectively. More details on the model are given in Ref.~\cite{La98,La99,La00}. Note that, due to the dominant $s\overline{s}$ nature of the $\phi$ meson, this channel is dominated by the exchange of the Pomeron.

Such a picture fails to reproduce the energy variation of the cross section of $J/\psi$ meson photoproduction. The reason is that the large mass of the charmed quark sets the hard scale and prevents the exchanged gluons to reinteract and form a Pomeron. Indeed, models based on the exchange of two gluons~\cite{RyXX,BrXX} are able to relate the rapid rise of the cross section with energy to the evolution of the gluon distribution in the proton. This finding was generalized to the description of exclusive vector meson electroproduction, at high virtuality $Q^2$, in terms of  Generalized Parton Distributions wich are discussed elsewhere~\cite{Gui01} in these proceedings.   

In this talk, I will explore another direction: I will increase the mometum transfer $t=(k_{\gamma}-k_V)^2$ to resolve the Pomeron and Reggeons, which are exchanged in exclusive photoproduction of vector mesons, into their simplest gluon or quark content.

\section{$\phi$ MESON PHOTOPRODUCTION}

\begin{figure}[h]
\centerline{\epsfig{file=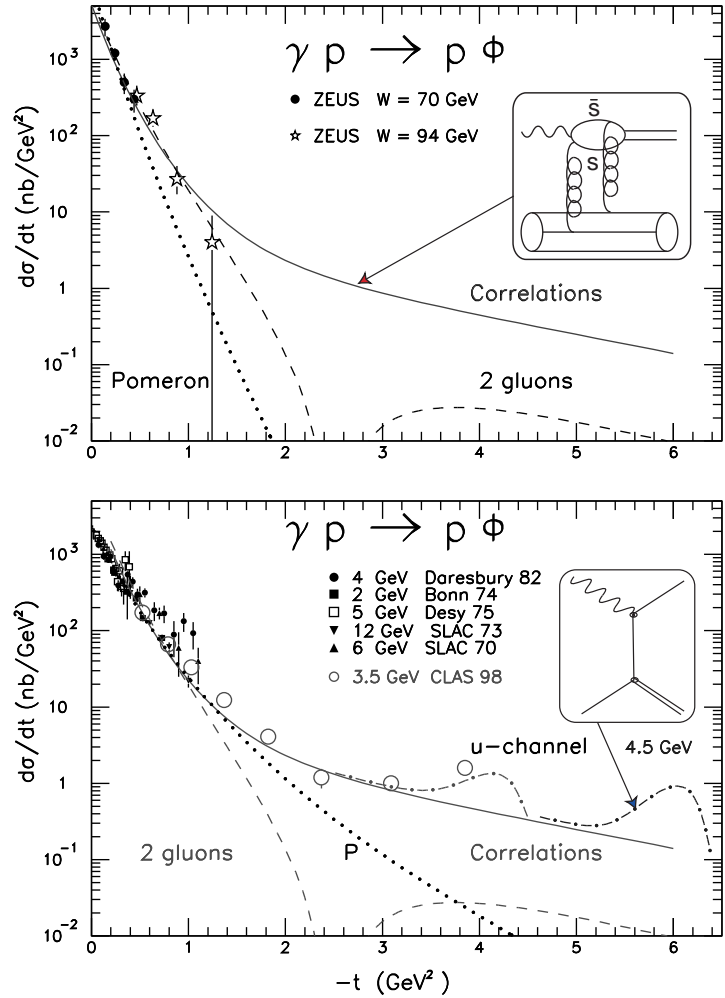,height=5.5in}}
\vspace{-1cm}
\caption{The differential cross section of the $p(\gamma, \phi)p$ reaction at HERA (top panel) and JLab (bottom panel). See ref.~\cite{La00}, for details.}
\label{phi}
\end{figure}

$\phi$ meson photoproduction allows to prepare a $s\bar s$ pair of strange quarks and study its interaction with hadronic matter. At low momentum transfer $t$ (small angle), its diffractive scattering is mediated by the exchange of the Pomeron. At high momentum transfer (large angle), the impact parameter is small and comparable to the gluon correlation length (the distance over which a gluon propagates before hadronizing): the Pomeron is resolved into its simplest component, two gluons which may couple to each of the quarks in the emitted vector meson or in the proton target. This is illustrated in Fig.~\ref{phi}, which shows data recently recorded at DESY~\cite{Br00} and JLab~\cite{An00}. At low $t$, the data confirm the shrinkage of the forward diffraction peak and the slow rise of the cross section with the energy, as expected from the exchange of the Pomeron Regge trajectory. The two gluon exchange contribution matches the Pomeron exchange contribution around $-t\sim 1$GeV$^2$ and reproduces the data at higher $t$. In the JLab energy range, $u$-channel nucleon exchange ``pollutes'' the highest $t$ bin: here the $\phi NN$ coupling constant $g_{\phi NN}= 3$  is the same as in the analysis of the nucleon electromagnetic form factors~\cite{Ja89}. More details may be found in ref.~\cite{La00}.

Such a finding is important as it tells us that, in the intermediate range of momentum  transfer  (let's say $1\leq -t\leq 10$ GeV~$^2$), large angle exclusive meson production can be understood in a perturbative way at the level of effective parton degrees of freedom: dressed quark and gluon propagators, constituent quark wave functions of the nucleon and of the meson, etc\ldots. At low momentum transfer (up to $-t\approx 2$~GeV$^2$), the cross section is driven by their integral properties: any nucleon wave function which reproduces the nucleon form factor leads to the same result; any gluon dressed propagator which reproduces the gluon correlation length ($\sim 0.2$ fm) leads to the same result. At higher momentum transfer, the cross section becomes more sensitive to the details of the wave function (giving access the quark correlations) and the shape of the gluon propagator. I refer to the contribution of F.~Cano~\cite{Ca01} for a more detailed account. Suffice to say that the large momentum cross section is reduced when either a more realistic wave function (than used in ref.~\cite{La00}), or a gluon propagator with a running mass (instead of a gaussian gluon propagator as in ref.~\cite{La00}), is used. 

At JLab, the experiment has been repeated at higher energy (around E$_{\gamma}$ = 4.5~GeV). The $u$-channel backward peak is moved at higher values of $-t$, leaving more room to reveal and check the two gluon exchange contribution. The preliminary results (not shown) confirm these predictions. Such studies provide us with a bridge with Lattice QCD calculations. The mass of  the gluon, $m_g(0)\approx 400$ Mev at vanishing virtuality, is close to the prediction of those calculations. In the near future they may also provide us with an estimate of quark correlations in the wave functions of hadrons.


\section{$\rho$ AND $\omega$  MESON PHOTOPRODUCTION}

\begin{figure}[h]
\centerline{\epsfig{file=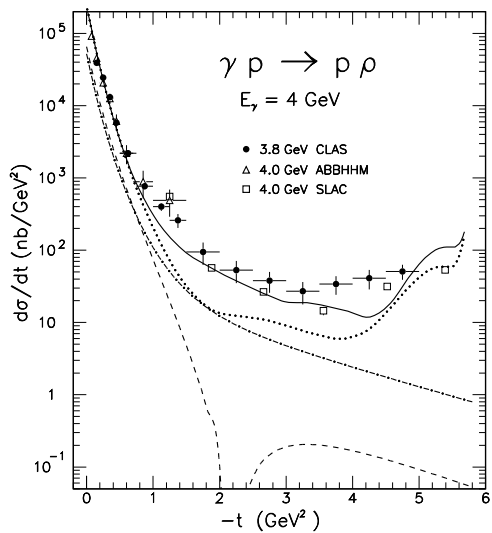,height=2.75in}}
\vspace{-1cm}
\caption{The differential cross section of the $p(\gamma, \rho)p$reaction, in the  JLab energy range. Dot-dashed (dashed) curves: two gluon exchange with (without) correlations. Dotted (full) curve: full model with linear (saturating) Regge trajectories.}
\label{rho_slac}
\end{figure}

In contrast to the $\phi$ meson sector, quark interchanges are not forbidden in the $\rho$ and $\omega$ meson photoproduction sector. Fig.~\ref{rho_slac} shows the latest data~\cite{Ba01} obtained at JLab. At low $-t$ the good agreement with the data is obtained adding, on top of the two gluon exchange amplitude, the exchange of the $f_2$ and $\sigma$ Regge trajectories, while the rise at high $-t$ comes from the $u$-channel exchange of the $N$  and $\Delta$ Regge trajectories (see ref.~\cite{La00}). At intermediate $-t$, the two gluon exchange contribution misses the data by a factor three. The agreement is restored when a saturating trajectory ($\alpha(t)=-1$ when $-t>3$ GeV$^2$) is used instead of a linear trajectory, for mesons. It leads to the asymptotic quark counting rules, which model-independently determine the energy dependency of the cross section at large -t. In a model where  the mesons are built from a potential, the linear confining potential leads to the linear long range part of the trajectory, which corresponds to the spectrum of the excited states of the meson, and the one gluon exchange short range potential leads to the saturating part of the trajectory~\cite{Ser94}.  In that sense, the exchange of the linear part of the trajectory can be viewed as the exchange of two quarks strongly interacting in a non pertubative way, while the exchange of the saturating part can be viewed as the  exchange of two quarks  interacting by the exchange of a single perturbative hard gluon. Saturating trajectories provides us with an economic and effective way to incorporate hard scattering mechanisms~\cite{Co84,Gui97}. 

\begin{figure}[h]
\centerline{\epsfig{file=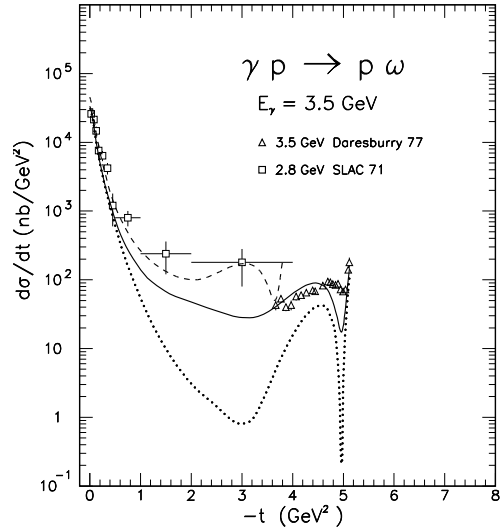,height=2.75in}}
\vspace{-1cm}
\caption{The differential cross section of the $p(\gamma, \omega )p$ reaction, in the  JLab energy range.  Dotted line: linear Regge trajectories at 3.5 GeV. Full (dashed) line: saturating Regge trajectories at 3.5 (2.8) GeV.}
\label{omega_kroll}
\end{figure}

Such a model reproduces also the sparse data in the $\omega$ production channel (Fig.~\ref{omega_kroll}). In this channel pion exchange dominates the cross section in the JLab energy range (see Fig.~\ref{vect_tot}), and the pion saturating trajetory is the same as the one which reproduces the cross section of the $\gamma p\rightarrow \pi^+ n$ reaction around $90^{\circ}$~\cite{Gui97}. The effect is more dramatic than in the $\rho$ channel, and I am eagerly waiting for the final analysis of the recent JLab data.

Note that the dip at hight $-t$ is due to the zero in the nucleon non degenerated Regge trajectory~\cite{Gui97}: this is the only trajectory which can be exchanged, contrary to the $\rho$ channel. 

\begin{figure}[p]
\centerline{\epsfig{file=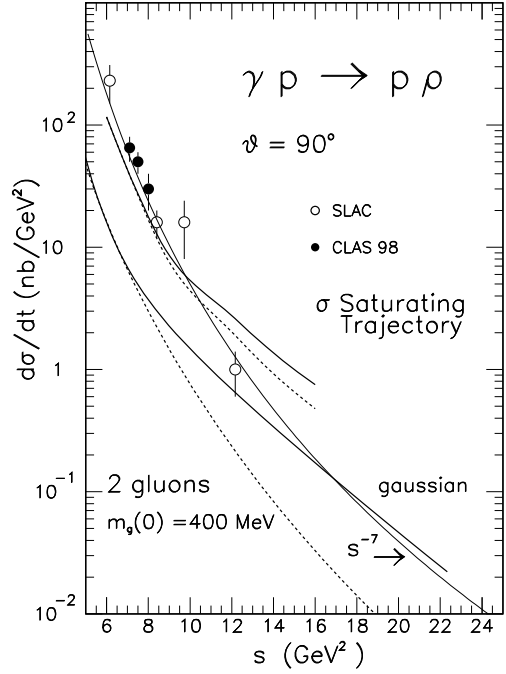,height=3.5in}}
\vspace{-1cm}
\caption{The $\theta =90^{\circ}$ cross section of the $p(\gamma, \rho)p$ reaction, in the  JLab/SLAC energy range. Bottom curves: two gluon exchange only. Top curves: full model. Dashed (full) curves: massive (gaussian) gluon  propagator.}
\label{rho_90}
\end{figure}

Such a model leads also to the $s^{-7}$ asymptotic behavior of the $\theta =90^{\circ}$ cross section of the $p(\gamma, \rho)p$ reaction depicted in Fig.~\ref{rho_90}. The two gluon exchange is too low, and the use of a massive gluon propagator ($m_g(0)=400$ MeV), instead of a gaussian one, decreases further its contribution at large $-t$. The bulk of the cross section comes from the exchange of the $\sigma$ Regge saturated trajectory.

\section{COMPTON SCATTERING}

At high energy, the real photon emitted in Compton scattering has enough time to fluctuate into a vector meson. Therefore, the Compton scattering cross section can be obtained by mutiplying the $\rho$ meson photoproduction cross section, which dominate over the $\omega$ and $\phi $ ones, by $4\pi \alpha_{em}/f_V^2$ (where $f_V$ is the radiative decay constant of the vector meson).

\begin{figure}[p]
\centerline{\epsfig{file=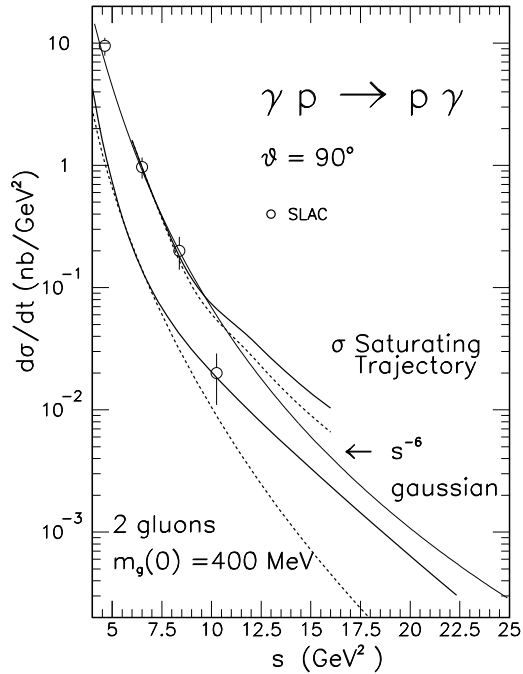,height=3.5in}}
\vspace{-1cm}
\caption{The $\theta =90^{\circ}$ cross section of the $p(\gamma, \gamma)p$ reaction, in the  JLab/SLAC energy range. Same meaning of the curves as in Fig.~\ref{rho_90}.}
\label{gam_90}
\end{figure}

\begin{figure}[h]
\centerline{\epsfig{file=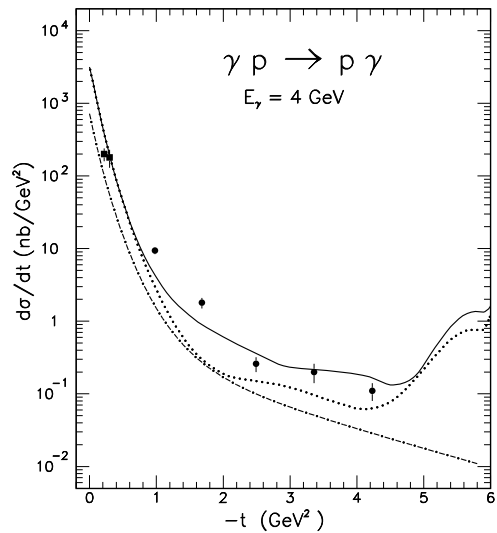,height=2.75in}}
\vspace{-1cm}
\caption{The differential cross section of the $p(\gamma, \gamma )p$ reaction at $E_{\gamma}=$   4~GeV. Same meaning of the curves as in Fig.~\ref{rho_slac}}
\label{gam_kroll}
\end{figure}

This leads to a good accounting of the sparse existing data: Fig.~\ref{gam_90} shows the excitation function at $\theta =90^{\circ}$, while Fig.~\ref{gam_kroll} shows the angular distribution at $E_{\gamma}=4$ GeV. A more comprehensive set of data will soon come from JLab.

\section{CHARM PHOTOPRODUCTION}

The threshold regime of charmonium production can provide a new window into multi-quark, gluonic, and hidden color correlations in hadronic wave functions. In contrast to charm production at high energy, charm production near threshold requires all of the target's constituents to act coherently in the  heavy quark production: only compact Fock states with radius of the order of the Compton wavelength of the charm quark can contribute to charm production at threshold.
Under these circumstances, one can relate~\cite{Bro01} the energy dependency of the cross section to the  behavior of its different gluonic components near $x=1$: $(1-x)^2$ for two gluon exchange, $(1-x)^0$ for three gluon exchange.

\begin{figure}[h]
\centerline{\epsfig{file=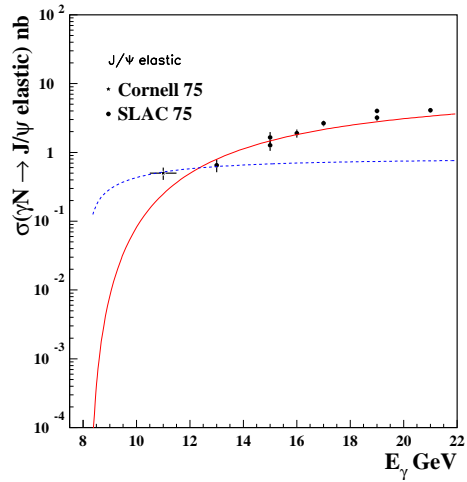,height=2.5in}}
\vspace{-1cm}
\caption{The threshold cross section of the $p(\gamma, J/\Psi  )p$ reaction. Solid line: two gluons. Dashed line: three gluons.}
\label{charm}
\end{figure}

Fig.~\ref{charm} shows that such an expectation is consistent with the sparse existing data and calls for a comprehensive dedicated measurement when the energy upgrade of JLab becomes a reality.

\section{CONCLUSIONS}

A consistent picture is emerging from the study of exclusive photoproduction of vector meson and Compton scattering.

At low momentum transfer, it relies on diffractive scattering of the hadronic contents of the photon, in a wide energy range from threshold up to the HERA energy domain.

At high momentum transfer, it relies on a partonic description of hard scattering mechanisms which provides us with a bridge with Lattice Gauge calculations. The dressed gluon and quark propagators have already  been estimated on lattice. One may expect that correlated constituent quark wave functions (at least their first moments) will soon be available from lattice. 

To day, JLab is the only laboratory which allows to explore this regime, thanks to its high luminosity.
Its current operation, at 4--6 GeV, has already revealed a few jewels. Its energy uprade to 12 GeV will permit a more comprehensive study of this field. Among other topics, this will allow to study the onset of asymptotic freedom, to determine the correlations between quarks and to single out the van der Walls part of the force between hadrons.  


\end{document}